\documentclass{iau} 
\usepackage{graphicx}
\usepackage{natbib} 
\usepackage{multicol}

\title[AGN and development in astronomy in Ethiopia and East-Africa] 
{Development in astronomy in Ethiopia and East-Africa through nuclear activity in galaxies}

\author[Mirjana Povi\'c]   
{Mirjana Povi\'c$^{1, 2}$}

\affiliation{$^1$Astronomy and Astrophysics Research and Develoment Division, Etiopian Space Science and Technology Institute (ESSTI), Addis Ababa, Ethiopia\\$^2$Instituto de Astrof\'isica de Andaluc\'ia (IAA-CSIC), Granada, Spain\\}

\pubyear{2019}
\volume{356}  
\jname{Nuclear Activity in Galaxies Across Cosmic Time}
\editors{M. Povi\'c, P. Marziani, J. Masegosa, H. Netzer,\\ S. H. Negu, \&
	S. B. Tessema, eds.}

\begin{document}

\maketitle

\begin{abstract}
In this paper we summarise the research that is currently going on in Ethiopia and East-Africa in extragalactic astronomy and physics of active galaxies and active galactic nuclei (AGN). The study is focused on some of the still open questions such as: what are the stellar ages and populations of ultra hard X-ray detected AGN and connection between AGN and their host galaxies?, what are the properties of AGN in galaxy clusters and the role that environment has in triggering nuclear activity?, what are the morphological properties of AGN and how precisely we can deal with morphological classification of active galaxies?, what are the properties of galaxies in the green valley and the role of AGN in galaxy evolution?, and what are the properties of radio-loud and radio-quiet quasars (QSO) and dichotomy between the two?. Each of these questions has been developed under one specific project that will be briefly introduced. These projects involve 6 PhD and 3 MSc students and collaborations between Ethiopia, Rwanda, South Africa, Uganda, Tanzania, Spain, Italy, and Chile. With all projects we aim: first, to contribute to our general knowledge about AGN, and second, to contribute to the development in astronomy and science in Ethiopia and East-Africa.

\keywords{galaxies: active; galaxies: main properties; astronomy development in Africa}
\end{abstract}

\firstsection 
              
\section{Introduction} 

How galaxies form and evolve is still one of the fundamental questions in modern cosmology. Many properties of different types of galaxies are still fairly understood, especially when going to higher redshifts. In particular, active galaxies, having active galactic nuclei (AGN) in their centre, play an important role in understanding galaxy formation and evolution, supermassive black hole (SMBH) formation and growth, star formation (SF) in galaxies, morphological transformation and role that AGN may have in moving galaxies from late- to early-types, or how the Universe was in its early stage \citep[e.g.,][]{heckman2014, netzer2015, hickox2018}. In the following we describe briefly several projects that are related with some of the still open questions in the field of AGN, and in the same time are important for human capacity building and astronomy development in Ethiopia and East-Africa.

\vspace{-0.6cm}

\section{Stellar ages and morphologies of ultra-hard AGN}
Connection between SF and AGN activity was studied widely over the past years, and shown to be very important for understanding the role of AGN in galaxy evolution \citep[e.g.,][]{povic2013, shimizu2017, masoura2018}. What are the stellar ages and average stellar populations of AGN host galaxies, and if there are differences depending on AGN type, are still some of open questions. The AGN sample detected in the ultra-hard X-rays (14\,-\,195\,keV) by the Swift BAT telescope is not affected by obscuration nor it is contaminated by stellar emission, and therefore presents some of the most unbiased samples. Therefore, the Swift-BAT AGN Spectroscopic Survey \citep[BASS\footnote{https://www.bass-survey.com/};][]{koss2017} gives us an unique opportunity to understand connection between AGN and their host galaxies by studying the SMBH mass and accretion rate \citep{koss2017, ricci2017} in relation to stellar properties, metallicities, and morphologies of AGN hosts. It is for the first time that this kind of study will be carried out for a complete sample of ultra-hard X-ray detected AGN. \\
\indent Using the optical spectra and BASS DR1 data \citep{koss2017} we are carrying out spectral energy distribution (SED) fitting by using the STARLIGHT code \citep{cidfernandes2004}, with aim to study stellar ages and populations of ultra-hard X-ray AGN. For fitting the type-2 AGN, we are following the same procedure as in \citep{povic2016}, and testing templates with different metallicities and stellar ages \cite{bruzual2003}. Using the obtained emission spectra we are measuring properties of all emission lines (integrated fluxes and equivalent widths), again using the same methodology as in \citep{povic2016}. Metallicities will be measured using the new Bayesian-like approach of \cite{PerezMontero2019} that has been tested on type-2 AGN. This study will be combined by morphological analysis of the same sample. Visual multiwavelength classification in optical, radio, and X-rays was carried out \citep[see][poster paper in Chapter 8]{Bilata2020}, where we obtained that most of ultra hard X-ray detected AGN are hosted by spirals in optical, are radio-quiet, and have compact morphologies in X-rays. We are finally planning to study in more details how/if X-ray luminosities, SMBH masses, and accretion rates are correlated with obtained stellar ages and populations, metallicities, and morphology, for understanding better connection between the ultra-hard AGN and their host galaxies. \\
\indent This project is carried out as a collaboration between Ethiopia and Spain. Morphological analysis resulted in MSc degree of Betelehem Bilata, one of our few female students.  

\vspace{-0.6cm}

\section{Properties of galaxies in galaxy clusters up to z\,$\sim$\,1.0}
The study of properties of galaxies inside galaxy clusters represents one of the main steps in understanding the formation and evolution of the Universe. In particular, it is important to understand how galaxies transform inside the clusters and how they change their properties as a function of redshift and environment. The research case proposed here has been carried out at the ESSTI under the GLACE collaboration \citep[GaLAxy Cluster Evolution survey;][]{SanchezPortal2015}, with general aim to study the properties and evolution of galaxies in galaxy clusters up to z\,$\sim$\,1. Our main objective is to better understand some of the still open questions such as: the role of AGN in galaxy clusters, metallicity variability, and galaxy transformation and evolution within clusters at different cosmic times. \\
\indent Two GLACE clusters, RXJ1257.2+4738 at z\,=\,0.866 and ZwCl0024.0+1652 at z\,=\,0.395, have been analysed using tunable filters (TF) data available from the OSIRIS instrument at the GTC 10m telescope and public data. For the RXJ1257 cluster we carried out morphological classification and analysis \citep{PintosCastro2016} using the non-parametric methods based on the galSVM code \citep{HuertasCompany2008}. Regarding ZwCl0024 cluster, emission line galaxies have been previously selected and analysed in H$_{\alpha}$ and [NII] lines \citep{SanchezPortal2015}. We carried out morphological classification of galaxies in ZwCl0024, in a consistent way as in RXJ1257, by classifying all galaxies as early- or late-types, and obtaining the most detailed catalogue up to date up to a clustercentric distance of 1\,Mpc \citep{BeyoroAmado2019}. Data reduction and analysis of H$_{\beta}$ and [OIII] lines in the ZwCl0024 cluster has been finalised. Obtained emission line fluxes and luminosities have been tested and compared with those of H$_{\alpha}$ and [NII] (Beyoro-Amado et al. 2020, in prep). We obtained pseudo-spectra of both lines and selected possible emission line galaxies after inspection of pseudo-spectra and HST/ACS images. We are studying the nuclear activity in this cluster using the BPT diagram and 4 emission lines from TF observations. Metallicity estimates will be carried out using [NII] line, SFR using H$_{\alpha}$ line, and extinctions using H$_{\alpha}$ and H$_{\beta}$ lines. All properties including morphology will be analysed in relation to local density and clustercentric distance. Finally, we will provide a more global analysis where properties measured for RXJ1257 and ZwCl0024 clusters will be compared with those obtained for local Virgo cluster using public data. For more information about this work see \cite{Beyoro2020} paper in Chapter 4.\\
\indent This project is related to PhD thesis of Zeleke Beyoro-Amado, and forms a part of collaboration between Ethiopia and Spain.  

\vspace{-0.6cm}

\section{Morphological properties of active galaxies}
Morphology is the most accessible indicator of galaxy physical structure, being crucial for understanding the formation of galaxies throughout cosmic time and for providing answers to some of still open questions mentioned above. In case of active galaxies there are still many inconsistencies between the results obtained and their interpretation regarding morphology \citep[e.g.,][]{pierce2007, georgakakis2008, gabor2009, povic2009, povic2012, mahoro2019}, and how the AGN affects morphological classification of its host galaxies \citep{gabor2009, cardamone2010, pierce2010}. In particular, the interpretation of morphology still remains a problem in the framework of galaxy evolution. Since measured morphology depends strongly on image resolution, classification in deep surveys remains very difficult, especially when dealing with faint and high redshifts sources \citep{povic2015}. Still detailed morphological analysis is missing in the field of AGN, especially at higher redshifts. \\
\indent In this work we are carrying out a detailed study on how the AGN contribution affects morphological classification of active galaxies. To do so, we are applying a similar method as used in \cite{povic2015}. Using a local sample of $\sim$\,2000 visually classified SDSS inactive galaxies \citep{nair2010} we simulated their images by adding in the centre different AGN contributions from 5\% up to 75\% of the total flux. By running the galSVM code \citep{HuertasCompany2008}, we are measuring six commonly used morphological parameters and are evaluating how well the morphological classification of active galaxies can be determined at z\,$\sim$\,0 and at higher redshift in COSMOS conditions (Getachew et al. 2020a, b, in prep.). Secondly, we are planning to estimate the possible AGN contribution of COSMOS galaxies by using GALFIT code \citep{Peng2002}. Finally, taking into account the previous results we will re-classify active galaxies in the COSMOS field and understand better how they evolved across cosmic time. For more information see \cite{GetachewWoreta2020} paper in Chapter 4.\\
\indent This project is related to PhD thesis of Tilahun Getachew-Woreta, and is carried out between Ethiopia and Spain. 

\vspace{-0.5cm}

\section{Properties of green valley galaxies and role of AGN in galaxy evolution}
On the colour-magnitude or colour-stellar mass diagrams green valley (GV) galaxies are located between the 'red sequence' and 'blue cloud' which are mainly populated by early- and late-type galaxies, respectively \citep[e.g.,][]{povic2013, schawinski2014, Salim2014, Bremer2018}. Different studies suggested that morphological transformation of galaxies happens in the GV during different timescales \citep{schawinski2014, Smethurst2015, Trayford2016, Bremer2018}, being also dependent on morphology \citep{NogueiraCavalcante2018}. Study of GV galaxies is therefore crucial for understanding the process of SF quenching, how galaxies transform from late- to early-types, and what is the role of AGN in morphological transformation of galaxies and their evolution. \\
\indent To understand better the properties of GV sources and connection between inactive and active galaxies we selected a large sample of sources and studied their properties such as SFRs, stellar masses, sizes, morphologies, stellar populations, and ages. We are carrying out this study at low, intermediate, and higher redshifts using public SDSS, GAMA, and COSMOS data, respectively. In addition, for a smaller number of sources we are using our own spectroscopic data from SALT telescope. We measured SFRs in COSMOS through the FIR Herschel/PACS data and SED fitting, and we observed the location of active and inactive galaxies on the main-sequence (MS) of SF. We found that most of our GV X-ray detected AGN with far-IR emission have SFRs higher than the ones of inactive galaxies at fixed stellar mass ranges. Therefore, they do not show signs of SF quenching, as shown in most of previous optical studies \citep[e.g.,][]{Nandra2007, povic2012, Leslie2016}, but rather its enhancement. Our results may suggest that for X-ray detected AGN with FIR emission if there is an influence of AGN feedback on SF in the GV the scenario of AGN positive feedback seems to take place, rather than the negative one \citep{mahoro2017}. In the second paper published recently, we studied morphological properties and found that a significant number of our AGN (38\%), but not the majority, are related to interactions and mergers \citep{mahoro2019}. At low redshifts using SDSS optical and GALEX UV data we analysed six criteria commonly used for selecting GV galaxies. We observed that depending on criteria different population of galaxies are selected, bringing therefore to different GV results (Nyiransengiyumva et al. 2020, in prep.). For understanding better the role of AGN in SF quenching in the GV, we are now measuring stellar ages, average stellar populations, and metallicities of selected active and inactive galaxies at both lower and higher redshifts. For more information see papers of \cite{mahoro2020} and \cite{beatrice2020} in Chapter 4.\\
\indent This project is related to PhD dissertations of Antoine Mahoro and Betrice Nyiransengiyumva from Rwanda, and is carried out as a collaboration between Rwanda, Ethiopia, South Africa, Uganda, and Spain.

\vspace{-0.5cm}

\section{Dichotomy of radio-loud and radio-quiet quasars}
Quasars (QSOs) were discovered more than 50 years ago. Being some of the most luminous sources in the Universe they are fundamental for cosmological studies \citep[see paper of][in Chapter 2]{Marziani2020}. Most of the QSOs in the local universe are radio-quiet (RQ). It is still under debate if there is an evidence for a continuity or physical dichotomy between RL and RQ QSOs, what is the origin of the powerful relativistic jets, and the effect that they have on the surrounding medium. Previous studies suggested the existence of two Populations (A and B) of QSOs in the 4D Eigenvector 1 (4DE1) plane defined by the FWHM of H$\beta$ and the strength of the optical FeII blend at 4570\AA~normalised by the intensity of H$\beta$ line (RFe), and the existence of 'QSO main sequence' \citep{Sulentic2000, Shen2014, Marziani2018}. It has been shown that RL QSOs are strongly concentrated in the Population B \citep{Zamfir2008}. In this work we are analysing the dichotomy of RL and RQ QSOs and the effect of the relativistic radio jets on the gas in the broad line emission region.\\
\indent We are using our spectroscopic data from CAHA 3.5m and GTC 10m telescopes of $\sim$\,60 RL QSOs. We want to quantify broad emission lines differences between RL and RQ sources, exploiting larger and more complete samples of QSOs with spectral coverage in H$\beta$, FeII, MgII and CIV emission lines than it was done previously. We are using SPECFIT code to determine the main parameters of each component. This will allow us to verify whether the larger values of FWHM(H$\beta$) among RL sources can be due to orientation, and study the wind properties affecting the profiles of the CIV and MgII lines. Currently we are analysing a sample of 12 RL QSOs using our CAHA spectra. We will focus our comparisons on RQ and RL sources that have the same mass and L/L$_{Edd}$ ranges. For more information see \cite{Terefe2020} paper in Chapter 8.\\
\indent This project is related with PhD thesis of Shimeles Terefe from Ethiopia, and is carried out in collaboration between Ethiopia, Spain, and Italy.

\vspace{-0.5cm}

\section{Conclusions}
This paper describes some of the main projects that are running under the extragalactic astronomy group at the ESSTI. Several other works have been conducted related with: 'Properties of inside-out assembled galaxies at z\,$<$\,0.1' \citep[see][in Chapter 8]{Zewdie2020}, 'Testing the alternative method to measure the accretion rate in galaxies' \citep[][in Chapter 8]{Gaulle2020}, and 'Characterisation of LINERs and retired galaxies at z\,$<$\,0.1' \citep[][in Chapter 8]{Mazengo2020}. These three projects resulted in three MSc dissertations in Ethiopia and Tanzania. Beside contributing to our general knowledge about galaxies and AGN, all mentioned projects contributed to development in astronomy and science in Ethiopia, Rwanda, Uganda, South Africa, and Tanzania. We managed to give more visibility to astronomy in Africa, to contribute to the institutional development of ESSTI and partner institutions, and to strengthen international collaborations. We are contributing to human capacity building of our first MSc and PhD students in the field, and we are also inspiring many other young people who went through different trainings and education and outreach activities that have been organised in Ethiopia and East-Africa over the past recent years.    

\vspace{0.3cm}

\small{\textbf{Acknowledgements\\}}
\small{Support of the Ethiopian Space Science and Technology Institute (ESSTI) under the Ethiopian Ministry of Innovation and
Technology (MInT) and Spanish MEC under AYA2016-76682-C3-1-P are gratefully acknowledged. This proceedings paper would not be possible without the IAUS 356 support of the IAU, ESSTI, EORC, MInT, ISP, IAA-CSIC, SEA, STFC-UKRI, DARA, ESSS, EA-ROAD, AAU, and Nature Astronomy.}

\begin{multicols}{2}

\end{multicols}

\end{document}